\documentclass[utf8]{frontiersSCNS} 
\usepackage{url,hyperref,lineno,microtype,subcaption}
\usepackage[onehalfspacing]{setspace}

\def\mgii{Mg{\sc ii}$\lambda$2800\/}
\def\apj{Astroph. J.}
\def\aj{Astron. J.}
\def\apjl{Astroph. J. Lett.}\def\apjs{Astroph. J. Supp.}
\def\mnras{Mon. Not. R. Astron. Soc.}
\def\aap{Astron. Astroph. }
\def\pasj{Publ. Astron. Soc. Japan}
\def\nat{Nature}\def\araa{Ann. Rev. Astron. Astroph.}
\defcitealias{marzianietal16}{M16}
\defcitealias{marzianietal16a}{M16a}
\def\mbh{$M_\mathrm{BH}$}

\def\kms{km s$^{-1}$\/}
\def\hb{H$\beta$}

\def\rfe{$R_\mathrm{FeII}$\/}
\def\lledd{$L/L_\mathrm{Edd}$}

\def\aliii{Al{\sc iii}$\lambda$1860\/}
\def\ciii{C{\sc iii}$\lambda$1909\/}
\def\siiii{Si{\sc iii}]$\lambda$1892\/}
\def\civ{C{\sc iv}$\lambda$1549\/}

\def\oiii{[O{\sc iii}]$\lambda$5007}
\def\feiiq{Fe{\sc ii}$\lambda$4570}
\def\oiiiopt{[O{\sc iii}]$\lambda\lambda$4959,5007}
\def\feii{Fe{\sc ii}}
\def\civonly{C{\sc iv}}
\definecolor{darkorange}{rgb}{1,0.612,0}
\definecolor{aquamarine}{rgb}{0.498,1,0.8314}

\newcolumntype{C}[1]{>{\centering\let\newline\\\arraybackslash\hspace{0pt}}m{#1}}
\def\keyFont{\fontsize{8}{11}\helveticabold }
\def\firstAuthorLast{Dultzin {et~al.}} 
\def\Authors{
D. Dultzin\,$^{1}$, P. Marziani\,$^{2,*}$,   J. A. de Diego\,$^{1}$,   C. A. Negrete\,$^{1}$, A. Del Olmo\,$^{3}$, M. L. Mart\'{i}nez-Aldama\,$^{3}$,  M. D'Onofrio\,$^{4}$,  E. Bon\,$^{5}$, N. Bon\,$^{5}$, G. M. Stirpe\,$^{6}$}
\def\Address{
$^{1}$\ Instituto de Astronom\'{i}a, UNAM, Mexico\\
$^{2}$ INAF, Osservatorio Astronomico di Padova, Italy \\
$^{3}$\ Instituto de Astrof\'{i}sica de Andaluc\'{i}a (IAA-CSIC), Granada, Spain\\ 
$^{4}$ Universit\`a di Padova, Padova, Italy\\  
$^{5}$Belgrade Astronomical Observatory, Belgrade, Serbia\\
$^{6}$INAF, Osservatorio di Astrofisica e Scienza dello Spazio di Bologna, Italy  }
\def\corrAuthor{Paola Marziani}
\def\corrEmail{paola.marziani@oapd.inaf.it}
\begin{document}
\onecolumn
\firstpage{1}
\title[ ]{Extreme quasars as distance indicators in cosmology} 
\author[\firstAuthorLast ]{\Authors} 
\address{} 
\correspondance{} 
\extraAuth{}
\maketitle
\begin{abstract}
Quasars accreting matter at very high rates (known as extreme Population A [xA] quasars, possibly associated with super-Eddington accreting massive black holes) may provide a new class of distance indicators covering cosmic epochs from present day up to less than 1 Gyr from the Big Bang.  At a more fundamental level, xA quasars are of special interest in studies of the physics of AGNs and host galaxy evolution. However, their observational properties are largely unknown. xA quasars can be identified in relatively large numbers from major optical surveys over a broad range of redshifts, and efficiently separated from other type-1 quasars thanks to selection criteria defined from the systematically-changing properties along the quasars main sequence. It has been possible to build a sample of { $\sim$ 250} quasars at low and intermediate redshift, and larger samples can be easily selected from the latest data releases of the Sloan Digital Sky Survey.   A large sample   can clarify the main properties of xA quasars which are expected  -- unlike the general population of quasars --  to radiate at an extreme, well defined Eddington ratio with small scatter.   As a result of the small scatter in Eddington ratio shown by xA quasars, we propose a method to derive the main cosmological parameters based on redshift-independent ``virial luminosity'' estimates from measurements of emission line widths, roughly equivalent to the luminosity estimates based from line width in early and late type galaxies. A major issue related to the cosmological application of the xA quasar luminosity estimates from line widths is the identification of proper emission lines whose broadening is predominantly virial over a wide range of  redshift and luminosity. We report on preliminary developments using the AlIII$\lambda$1860 intermediate ionization line and the Hydrogen Balmer line H$\beta$\ as virial broadening estimators, and we briefly discuss the perspective of the method based on xA quasars. 

\tiny
 \keyFont{ \section{Keywords:}   quasar main sequence -- line profiles --  emission lines  --  supermassive black holes -- black hole physics -- dark energy -- cosmological paremeters -- cosmology } 
\end{abstract}

\section{Introduction}
\label{intro}

Quasars show a rich spectrum of emission lines coming from a side range of ionic species \citep[e.g.,][]{netzer13}. Differences in line profiles, line shifts, line intensities  hint at {diversity} in dynamical conditions and ionization levels of the broad line region (BLR; \citealt{sulenticetal00a}).  There is a general consensus that the \hb\ and low-ionization lines (LILs) in general are mainly broadened via Doppler effect from emitting gas virial motion, while high-ionization lines (HILs)  are more affected by non-virial broadening, probably associated with a disk wind or clumpy outflows \citep[e.g.,][for a variety of approaches]{marzianietal96,progaetal00,elvis00,richardsetal11}. The relative prominence of the wind and disk component reflects the different balance between radiation and gravitation forces \citep{ferlandetal09,marzianietal10,netzermarziani10}, and is systematically changing along the quasar main sequence (MS). If the main driver of the stellar H-R diagram can be considered  mass, for the quasar main sequence it is the luminosity to black hole mass ratio that apparently matters most \citep{borosongreen92,sulenticetal00a,marzianietal01,shenho14,sunshen15,pandaetal19}. In addition, unlike stars,  quasars are not endowed of spherical symmetry but only by {\em axial symmetry}: the spin of the black hole and the axis of the accretion disk define axes of symmetry and a preferential plane (i.e., the equatorial plane of the black hole, the inner accretion disk plane). Aspect becomes an important parameters for AGN. The effect is most noticeable in  the distinction between type-1 and type-2 (obscured) AGN \citep[e.g., ][]{antonucci93,urrypadovani95}, but is relevant also for type-1 AGN (unobscured), as the viewing angle (defined as the angle between the disk axis and the line of sight) may range between 0 and 60 degrees.   

Quoting \citet{marzianietal01}, we can say that {\em \ldots the observed correlation \ldots  can be accounted for if it is primarily driven by the ratio of AGN luminosity to black hole mass ($L /M \propto\, $ \lledd, Eddington ratio) convolved with source orientation.}  Several recent reviews   papers have dealt with the quasar main sequence \citep[e.g.,][]{sulenticetal08,sulenticmarziani15,marzianietal18}.  Here we will summarize briefly the basic concepts and trends (\S \ref{ms}), afterward focusing on a particular class of objects, the strongest \feii\ emitters along the sequence (\S \ref{xA}). The basis of the virial luminosity estimations are then presented in \S \ref{virial}, also in an aspect-angle dependent form. Preliminary results for cosmology are illustrated in \S \ref{cosmo}.  

\section{A main sequence  for type-1 (unobscured) quasars}
\label{ms}

The Main Sequence stems from the definition  of the so-called  Eigenvector 1 of quasars   by a Principal Component Analysis of PG quasars \citep{borosongreen92}.  The MS of quasars is customarily presented in an optical plane defined by the parameter of \feii\ prominence \rfe\ { (the intensity ratio between the FeII blend at $\lambda$4570 and H$\beta$)} and by the { full-width half maximum} (FWHM) of \hb\ \citep{borosongreen92,sulenticetal00a}, and associated with the anti- correlation between strength of \feiiq\ (or \oiii\ peak intensity) and width of \hb.  \feii\  {maintains  the same relative intensity of the emission blends (at least to a first approximation)} but \feii\ intensity with respect to \hb\ changes from object to objects, and in the strongest \feii\ emitters, \feii\  emission can dominate the thermal balance of the BLR \citep{marinelloetal16}. Therefore it does not appear surprising that the \feii\ dominance parameters is a fundamental parameter in organising the diversity of type-1 quasars. The FWHM \hb\ is related to the velocity field in the low-ionisation part of the BLR, and ultimately associated with the { orientation } angle, black hole mass and Eddington ratio (\S \ref{popa}). Since 1992, the Eigenvector 1 MS has been found in increasingly larger samples, involving more and more extended sets of multifrequency parameters \citep{sulenticetal07,zamfiretal10,shenho14,wolfetal19}. In addition to the parameters defining the optical plane, \rfe\ and FWHM(\hb), several multifrequency parameters related to the accretion process and the accompanying outflows are also correlated \citep[see][for an exhaustive list]{fraix-burnetetal17}.

The MS allows for the definition of spectral types \citep{sulenticetal02,shenho14}. In addition \citet{sulenticetal00a} introduced  a break at 4000 \kms\ and the distinction between Population A (FWHM \hb$\lesssim$4000 \kms) which  includes the Narrow Line Seyfert 1 (NLSy1s), and Population B of broader sources. {  Extreme Population A (xA) sources are the ones toward the high \rfe\ end  of the main sequence (\rfe $>$ 1). } The optical spectra of the prototypical Narrow Line Seyfert 1 (NLSy1)  I Zw 1 and the Seyfert-1 nucleus  NGC 5548 illustrate the changes that occur along the sequence. I Zw 1  { (a prototypical xA source)} shows a low-ionization appearance, with strong \feii, weak \oiiiopt, and a spiky \hb\ broad profile; on the converse the optical spectrum of NGC 5548  suggests a higher degree of ionization, with low \feii\ emission, strong \oiiiopt. It is expedient to consider the Hydrogen Balmer line  \hb\ and the  \civ\ as prototypical optical and UV  LILs and HILs, respectively. Comparing the \civ\ and \hb\ lines yields important information: in the spectra of NGC 5548 the two lines show similar profile, while in the ones of I Zw 1 the \civ\ emission is almost fully blue shifted with respect to a predominantly symmetric and unshifted (with respect to rest frame) \hb\ profile \citep[e.g., ][]{sulenticetal00a,leighly04,coatmanetal16}. 

The comparison between LILs and HILs has provided insightful constraints of the BLR at low-$z$\ \citep{marzianietal96}, and this is even more true if the comparison of  \hb\ and \civ\ is carried out  at high $L$ \ \citep{sulenticetal17,bisognietal17,shen16,vietrietal18}.  Perhaps the most remarkable fact is that a LIL-BLR appears to remain basically virialized \citep{marzianietal09,sulenticetal17}. The \civ\ blueshift depends on luminosity   (the median $c(\frac{1}{2})$ is $\approx$\ 2600 \kms\ and 1800 \kms\ for Pop. A and B { radio-quiet} (RQ) at high-$L$ against less than 1000 \kms at low-$L$), but the dependence is not strong, and can be accounted for in the framework of a radiation driven winds.  In spite of the  extremely high amplitude \civ\ blueshifts of Pop. A quasars at the highest luminosity $L \gtrsim 10^{47}$  erg s$^{-1}$, the \hb\ profile remains (almost) symmetric and unshifted with respect to rest frame.  Fig. \ref{fig:virialhb} shows the emission line profiles of \hb, \civ, and \aliii\ and \siiii\ in three quasars of widely different luminosity. LILs in the optical and UV remains symmetric and unshifted. The lines have been decomposed into two main components. 

\begin{itemize}
\item  The broad component (BC), also known as   the intermediate  component, the core component or the central broad component  following various authors \citep[e.g.,][]{brothertonetal94a,popovicetal02,adhikarietal16,kovacevicdojcinovicpopopvic15}. Following \citet{sulenticetal02}, the BC is {modeled} by a symmetric and unshifted profile  (Lorentzian for  Pop. A or Gaussian for Pop. B), {as it is believed} to be associated with a virialized BLR subsystem. The { virialized} BLR could be defined as the {BLR} subregion that is in the condition to {emit FeII.} { Given the stratification of ionization in the BLR, the \feii\ emission is apparently weighted toward outer radii with respect to \hb\  \citep{barthetal13}, as also supported by the slightly narrower \feii\ with respect to \hb\ \citep[e.g., ][]{sulenticetal04}. The regions emitting \hb\ BC\footnote{Note that in Population B the \hb\ profile can be decomposed into a BC and a broader line base, the very broad component (VBC).  Attempts at fitting \feii\ emission with a VBC in Pop. B did not produce meaningful results.} and \feii\ are however believed to be largely co-spatial.}
  
\item The blue shifted component (BLUE).  A strong  blue excess in Pop. A \civonly\ profiles is {obvious, as in some \civonly\ profiles -- like the one of the xA  prototype I Zw1 -- BLUE  dominates} the total emission line flux \citep{marzianietal96,leighlymoore04}.  For BLUE, { there is no evidence of a ``stable'' profile, and a ``stable'' profile is not expected, as we are dealing with a component that is likely subject to phenomena associated with turbulence and hydrodynamic instabilities. In our recent work \citep{sulenticetal17}, we have  } modeled the blueshifted excess adding to the symmetric BC profile one or more skew-normal distributions   \citep{azzaliniregoli12}. The ``asymmetric Gaussian'' {fitting function} is, at present, empirically motivated  by the often-irregular  { BLUE} profile in \civonly\ and \mgii.  
\end{itemize}

The  line width of the LILs clearly increases with luminosity passing from $\sim 1000$ \kms\ to 5000 \kms\ from $\log L \sim$ 44 to $\sim 47$.  This behaviour — typical of Pop. A quasars  — suggests that   a virialized sub-system emitting mainly LILs coexists with outflowing gas.  Even at the highest luminosity the data \citep{sulenticetal17,vietrietal18,coatmanetal16} remain consistent with a scenario that posits a dichotomy within the BLR, where the blue shifted emission is produced in clumps of gas or a continuous wind  \citep{collin-souffrinetal88,elvis00}, and LILs are emitted by gas in virial motion. In this scenario, we observe a net blueshift in \civ\ because of the large gas velocity field component along the radial direction, with the receding side of the flow obscured by an optically thick equatorial structure, most likely associated with the accretion disk. I Zw 1 fits very well this interpretation: if LILs are emitted in the flattened disk structure, the \hb\ should appear narrow, while \civ\ emission broader and shifted to the blue, which is indeed the case \citep{marzianietal96,leighly04}.

\subsection{The definition of spectral types}

The data point occupation in the plane \rfe -- FWHM(\hb)   allows for the definition of spectral  NLSy1s types; A1, A2... in order of increasing \rfe; A1, B1, B1+ .\ldots in order of increasing FWHM(\hb). This has the considerable advantage that a composite over all spectra within each bin should be representative of objects in similar dynamical and physical conditions.  In principle, a prototypical object can be defined for each spectral type to analyze systematic changes along the quasar MS.  Before summarizing some basic trends, it is helpful to recall the main motivations behind the distinction between Population A and B.

\subsection{Population A and NLSy1s}
 \label{popa}

Why choose a FWHM limit for \hb\ at 4000 \kms?  Many studies even in recent times distinguish between NLSy1s (FWHM \hb\ $\lesssim 2000$ \kms) and rest of type 1 AGNs \citep[e.g.,][]{craccoetal16}.  The basic reason to extend the limit to 4000 \kms\ is that several important properties  of NLSy1s are not distinguishable from ``the rest of Population A'' in the range $2000$ \kms $\lesssim $ FWHM(\hb) $\lesssim 4000$ \kms.  NLSy1s are located at the low-end of the distribution of FWHM(H$\beta$) in the samples of \citet{shenetal11,zamfiretal10} without any apparent discontinuity.  NLSy1s and the rest of Population A  show no apparent discontinuity as far as line profiles are concerned \citep{craccoetal16}.  Composite H$\beta$\ profiles of spectral types along the MS are consistent with a Lorentzian as do the rest of Population A   sources. A recent  analysis  of composite line profiles in narrow FWHM(H$\beta$) ranges ($\delta $ FWHM = 1000 \kms) confirms that there is no discontinuity at 2000 \kms; best fits of  profiles remain Lorentzian up to at least 4000 \kms. A change in the \hb\ profile shape occurs around FWHM(H$\beta$)  = 4000 \kms \ at the Population A limit, not at the one of NLSy1s \citep{sulenticetal02}. No significant difference between CIV$\lambda$1549 centroids  at half-maximum of NLSy1s and rest of Pop. A is detected using the HST/FOS data of the \citet{sulenticetal07} sample \citep{marzianietal18a}.   In addition NLSy1s span a broad range of spectral types, the $R_\mathrm{FeII}$ parameter being between almost 0 and 2 in the quasi-totality of sources, as the rest of Population A   sources do.

{A complete tracing of the MS at high $L$\ is still missing} (we consider high-luminosity sources the quasars with bolometric $\log L \gtrsim 47$ [erg/s]): the \hb\ spectral range is accessible only with IR spectrometers, and high-luminosity quasars are exceedingly rare also at $z \lesssim 1$. 
The main effect of a systematic increase in black hole mass \mbh \ can be predicted.  If the motion in the LIL-BLR is predominantly virial, we can write the central black hole mass \mbh\ as a virial mass :

\begin{equation}
M_{\rm BH} = \frac{r_\mathrm{BLR} \delta v_{\rm K}^{2}}{G}
\label{eq:vmbh}
\end{equation}

which follows from the application of the virial theorem in case all gravitational potential is associated with a central mass. $\delta v_{\rm K}$ is the virial velocity module, $r_\mathrm{BLR}$\ the radius of the BLR, $G$\ the gravitational constant. Eq. \ref{eq:vmbh} can be of use if we can relate $\delta v_{\rm K}$\ to the observed velocity dispersion, represented here by the FWHM of the line profile: 

\begin{equation}
M_{\rm BH} = f_{\rm S} \frac{r  {\rm_\mathrm{BLR} FWHM}^{2}}{G}
\end{equation}

via the structure factor (a.k.a. as form or virial factor)    whose definition is given by:

\begin{equation}
 \delta v_{\rm K}^{2} = f_{\rm S}{\rm FWHM}^{2}
\end{equation}

If the BLR radius follows a scaling power-law with luminosity ($r \propto L^\mathrm{a}$, \citealt{kaspietal00,bentzetal13}), then 

 \begin{equation}
\mathrm{FWHM}   \propto  f_\mathrm{S}^{-\frac{1}{2}}  {L}^\frac{1-a}{2} \left(\frac{L}{M_\mathrm{BH}}\right)^{-\frac{1}{2} }
\end{equation}

Or, equivalently, 
 
\begin{equation}
\mathrm{FWHM}   \propto f_\mathrm{S}^{-\frac{1}{2}} \left(\frac{L}{M_\mathrm{BH}}\right)^{-\frac{a}{2}} M_\mathrm{BH}^\frac{1-a}{2}. 
\end{equation}

If $a = 0.5$, the FWHM grows with \mbh$^{0.25}$, meaning a factor of 10 (!) for $\log L$ passing from 44 (relatively low luminosity) to 48 (very luminous) quasars. The trend may not be detectable in low-$z$\ flux limited samples, but becomes appreciable if sources over a wide range in $L$ are considered.   At high \mbh, the MS becomes displaced toward higher FWHM values in the optical plane of the MS; the  displacement most likely accounts for the wedge shaped appearance of the MS if large samples of SDSS quasars are considered \citep{shenho14}. 

If we consider a physical criterion for the distinction between Pop. A and B a limiting Eddington ratio ($\log $ \lledd $\sim$ 0.1 — 0.2), then the separation at fixed \lledd\ becomes luminosity dependent. In addition FWHM(\hb) is strongly sensitive to viewing angle, via the dependence on $\theta$ of the $f_\mathrm{S}$\ (\S \ref{angle}).  An important consequence is that a strict FWHM limit has no clear physical meaning: its interpretation depends on sample properties. At low and moderate luminosity ($\log L \lesssim$ 47 [erg s$^{-1}$]), the limit at 4000 \kms\ selects mainly sources with $\log $ \lledd $\gtrsim$ 0.1 — 0.2 i.e., above a threshold that is also expected to be relevant in terms of accretion disk structure. However, the selection is not rigorous, since a minority of low \lledd\ sources observed pole-on (and hence with FWHM narrower because of a pure orientation effect) are expected to be below the threshold FWHM at 4000 \kms. Such sources include core-dominated radio-loud quasars whose viewing angle $\theta$\ should be relatively small \citep{marzianietal01,zamfiretal08}. They are expected to be rare, because the probability of observing a source at a viewing angle $\theta$\ is $P(\theta) \propto \sin \theta$, { although their numbers is increased in flux-limited sample because of the Malmquist effect}. As a conclusion, we can state that separating Pop. A and B at 4000 \kms\  makes sense for low $z$ sample and that, by a fortunate circumstances, Pop. A includes the wide majority of relatively high \lledd\ sources.

 
 \begin{figure}[h!]
\begin{center}
\includegraphics[width=15cm]{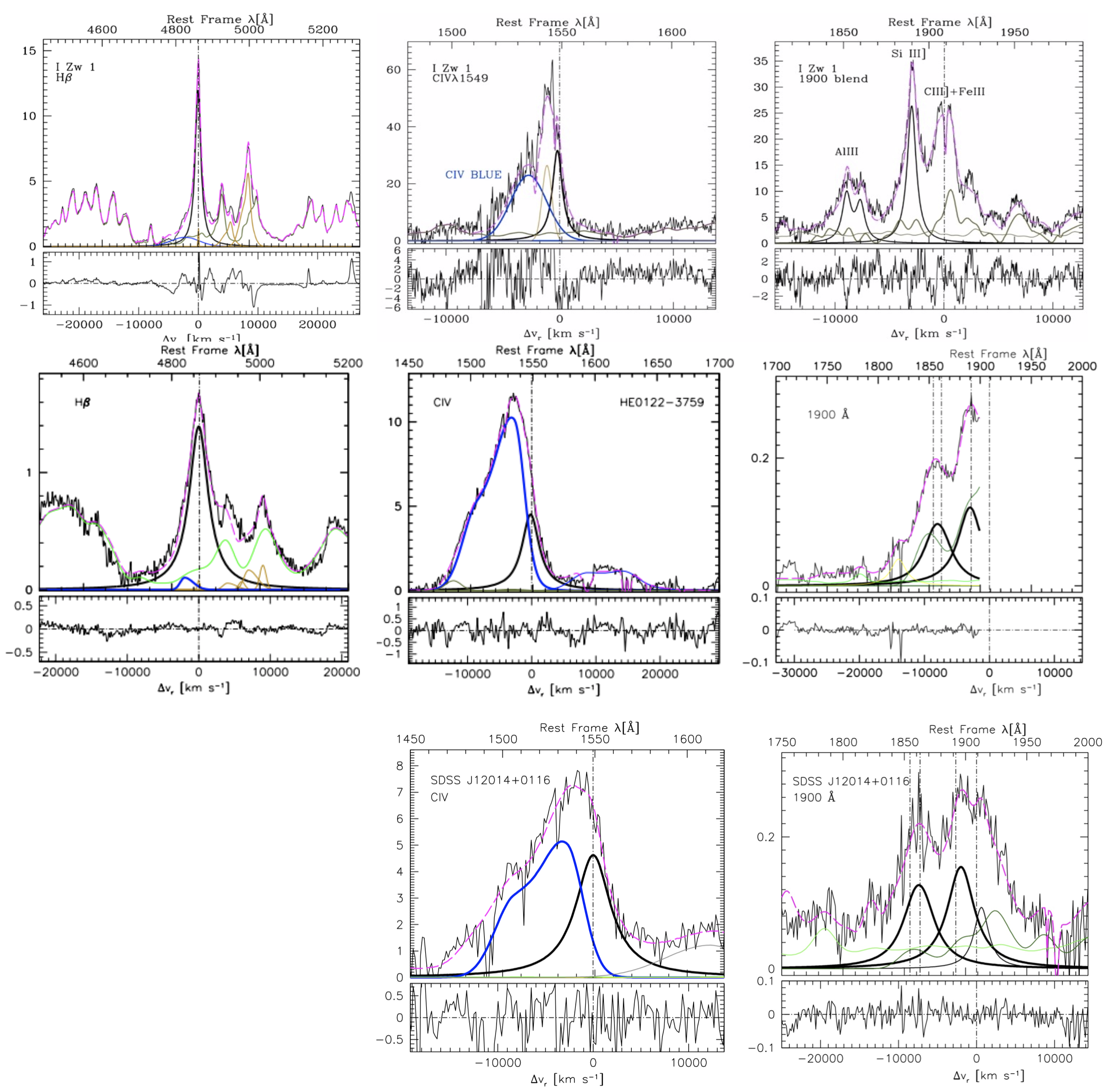}
\end{center}
\caption{Intercomparison of line profiles of \hb (left), \civ\ (middle) and \aliii\ in the 1900 blend for three xA quasars of widely different luminosity (top: I Zw 1 \citep{marzianietal10}; middle: HE0122-3759 \citep{sulenticetal17}; bottom: SDSS J12014-0116 \citep{negreteetal12}). The thin black line shows the continuum subtracted spectra, the thick black lines trace the virialized symmetric broad component, and the thick blue line the blueshifted excess. \feii\ emission is shown in lemon green, and narrow lines in orange color. The magenta dot-dashed line traces the model of the spectrum including all emission components. Note the absence or weakness of the BLUE for both \hb\ and \aliii.  }\label{fig:virialhb}
\end{figure}
 
 \section{Extreme Population A: Selection and physical conditions}
\label{xA} 
 
As mentioned in \S \ref{intro}, Eddington ratio is a parameter that is probably behind the \rfe\ sequence in the optical plane of the MS. It is still unclear why it is so. At a first glance, the low-ionisation appearance of sources whose Eddington ratio is expected to be higher is puzzling.   \citet{marzianietal01} accounted for this results including several observational trends that could lower the ionisation parameter with increasing \rfe. More recent results include a careful assessment of the role of metallicity \citep{pandaetal18,pandaetal19}. There is empirical evidence of a correlation between \rfe\ and \lledd, but the issue is compounded with the strong effect of orientation on the line broadening, and hence on \mbh\ and \lledd\ computation. As an estimate of the viewing angle is not possible for individual radio-quiet sources (even if some recent works provide new perspectives), conventional \mbh\ (Eq. \ref{eq:vmbh}) and \lledd\ estimates following the definitions in the previous section  suffer of a large amplitude scatter \citep[][and references therein]{marzianietal19}. Mass estimates over a broad range of FWHM are especially uncertain, and may lead to sample-dependent,  spurious trends.  Even if a combination of effects related to  Eddington ratio, viewing angle, and metal content reproduce the sequence, and accounts for the relative distribution of sources in each spectral bin, the actual structure of the BLR and its relation with the accretion mode is still unclear.  This said, if our basic interpretation of \rfe\ as mainly related to \lledd\ is correct, then the strongest \feii\ emitters should be the highest radiators. For instance, \citet{sunshen15}  provide evidence not dependent on the viewing angle:  in narrow bins of luminosity, the stellar velocity dispersion of the host spheroid (a proxy for \mbh) is anti correlated with \rfe, implying that \lledd\ increases with \feii. The fundamental plane of accreting black holes \citep{duetal16a} correlates \lledd\ and dimensionless accretion rate with \rfe\ and with the “Gaussianity” parameter $D = $FWHM/$\sigma$ of \hb. Both methods imply that the highest \rfe\ corresponds to the highest \lledd. 

If we select spectral types along the sequence satisfying the condition \rfe $\gtrsim  1$ (A3, A4, \ldots), we select spectra with distinctively strong \feii\ emission and Lorentzian Balmer line profiles. They account for $\sim$ 10\% of quasars in low-z, optically selected sample FeII in Pop. A. 


The simple selection criterion

 \begin{itemize}
\item  \rfe = FeII$\lambda$4570 blend/\hb\ $>$ 1.0
\end{itemize}

corresponds to the selection criterion:

 \begin{itemize}
\item   AlIII $\lambda$1860/SiIII]$\lambda$1892$>$0.5 \& SiIII]$\lambda$1892/ CIII]$\lambda$1909$>$1
\end{itemize}

The  \rfe$>$1 and the UV  selection criteria are believed to be equivalent. \citet{marzianisulentic14} compared the distribution of sources satisfying \rfe$\gtrsim 1$ in the plane AlIII $\lambda$1860/SiIII]$\lambda$1892 vs SiIII]$\lambda$1892/ CIII]$\lambda$1909, and found that these sources were confined in a box satisfying the boundary conditions AlIII $\lambda$1860/SiIII]$\lambda$1892$>$0.5 \& SiIII]$\lambda$1892/ CIII]$\lambda$1909$>$1, as expected if the two selection criteria are consistent. The number of sources for which this check could be carried out is however small, and xA sources are located in borderline positions. Work is in progress to test the consistency on a larger sample. 

However, the UV xA spectrum is very easily recognizable, as shown by the composite spectrum of \citet{martinez-aldamaetal18}.  Lines have low equivalent width: some xAs are weak lined quasars \citep[W(\civ) $\le$ 10 \AA, WLQ][]{diamond-stanicetal09}, whereas WLQs can be considered as the extreme of extreme Pop. A.  The \ciii\ emission almost disappears. In the plane $\log U - \log n_\mathrm{H}$\ defined by CLOUDY simulation, UV line intensity ratio converge toward extreme values for density (high, $n_\mathrm{H} > 10^{12}-10^{13}$ cm$^{3}$), ionization (low, ionization parameter $U\sim10^{-3} - 10^{-2.5}$).  Extreme values of metallicity  are also derived from the intensity ratios CIV/AlIII CIV/HeII AlIII/SiIII] \citep[][Sniegowska et al. 2019 in preparation]{negreteetal12,martinez-aldamaetal18}),  most likely above 20 times solar or with abundances anomalies that increase selectively aluminum or silicon, or both. 

\begin{figure}[htp!]
\begin{minipage}[t]{0.475\linewidth}
\centering
\hspace{-1.cm}
\includegraphics[width=8cm]{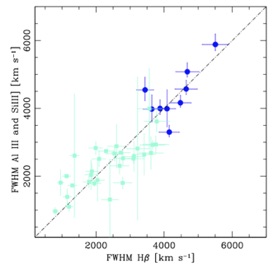}
\end{minipage}
\hspace{0.5cm}
\begin{minipage}[t]{0.475\linewidth}
\centering 
\vspace{-1.5in}
\caption{FWHM of \aliii\ vs FWHM \hb\ in \kms\ for Pop. A sources. The pale blue dots represent quasars from the RQ Pop. A sample of \citet{sulenticetal07}, while the blue are from the high-luminosity sample of \citet{sulenticetal17}. The dot-dashed line traces the 1:1 relation. \label{fig:hbal}
 }
\end{minipage}
\end{figure}

In addition to the selection criteria, a virial broadening estimator equivalent to \hb\ should be defined from the emission line in the rest frame UV spectrum. The \civ\ line is unsuitable unless heavy corrections are applied \citep{coatmanetal17,marzianietal19}.  However, the UV spectrum of xA quasars at z $\sim 2$ show symmetric low-ionization and blueshifed high- ionization lines even at the highest luminosity
\citep[][and reference therein]{martinez-aldamaetal18}. The results of a preliminary analysis is shown in Fig. \ref{fig:hbal}: the FWHM of \hb\ and \aliii\ for Pop. A objects are well correlated, with no systematic deviation from the 1:1 relation. Work is in progress to focus this systematic comparison to a sizeable sample of xA sources.   

The Eddington ratio precise values depend on the normalization applied; the relevant result is that xA quasars radiate at extreme \lledd\ along the MS with small dispersion \citep{marzianisulentic14}. This results is consistent from the expectation of accretion disks at very large accretion rates. Accretion disk theory { predicts} low radiative efficiency at high accretion rate and that \lledd\ saturates toward a limiting values  \citep{mineshigeetal00,abramowiczetal88,sadowskietal14}. { The \lledd\ distribution shown by \citet{marzianisulentic14} provides some support to this finding. While \rfe\ and \lledd\ might be  correlated, as mentioned above, if \rfe $\gtrsim 1$, \lledd\ apparently scatters around a well-defined value with a relatively small scatter $\approx 0.13 $\ dex. Clearly, this results should be tested by larger sample and, even more importantly, by Eddington ratio estimators not employing the FWHM of the line used for the \mbh\ computation.}  Another important fact is the self similarity of the spectra selected by the \rfe\ criterion: as Fig. \ref{fig:virialhb} shows, the LILs broadens from 1000 to over 5000 km/s, but the relative intensity ratios (and so the overly appearance of the spectrum) remains basically unchanged. This occur over an extremely wide luminosity range,  $\log L \sim 44 - 48$ [erg s$^{-1}$], covering local type 1 quasars in the local Universe as well as the most luminous quasars that are now extinct but that were shining bright at redshift 2 when the Universe had one quarter of its present age.

Accretion disk theory predicts that at high accretion rate  a geometrically thick, advection dominated disk should develop \citep{abramowiczetal88,sadowskietal14}, especially at the radii closest to the central black hole where the temperature should be so high that radiation pressure dominates over gas pressure. The disk vertical structure and the interplay between BLR and disk remains to be clarified \citep{wangetal14a}. It is one of the biggest challenges of the next decade. The issue can be addressed by two dimensional reverberation mapping and by careful modelling of the coupling between dynamical and physical conditions \citep{lietal13,pancoastetal14,lietal18}. A tentative model of xA sources is provided by the sketch of Fig. 6 of \citet{marzianietal18}. The innermost part of the disk is puffed up by radiation pressure, while the outermost one remains geometrically thin. This change from the standard thin alpha disk provides two key elements for the BLR structure: the existence of a collimated cone like region, where the high ionization outflows might be produced. While a vertical shape of the outer side of the geometrically thick region is unlikely to be appropriate, shadowing of the low-ionization emitting region is expected to occur, with the unpleasant consequence that the continuum seen by the BLR may not be the one seen by the observer \citep{wangetal14a}. 

\section{Eddington standard candles as distance indicators for cosmology	} 
\label{virial} 

From the discussion of the previous sections we gather that three conditions are satisfied for xA quasars: 
               
\begin{enumerate}
\item  constant Eddington ratio \lledd; xA quasars radiate close to Eddington limit:
\begin{equation}
\frac{L}{L_\mathrm{Edd}} = \lambda_\mathrm{E} \propto \frac{L}{M_\mathrm{BH}}
\end{equation}
Note that the precise value or \lledd\ depends on the normalisation applied for \mbh\ and on the bolometric correction to compute the bolometric luminosity. Applying widely employed scaling laws, \lledd $\lesssim 1$. 

\item The assumption of  virial motions of the low-ionization BLR, so that the black hole mass \mbh\ can be expressed by the virial relation (Eq. \ref{eq:vmbh}):

\begin{equation}
M_{\rm BH} = \frac{r_\mathrm{BLR} \delta v_{\rm K}^{2}}{G}\nonumber
\end{equation}

\item Spectral invariance: for extreme Population A, the ionization parameter $U$ can be written as 
\begin{equation}
U = {Q(H)}/{4 \pi r_\mathrm{BLR}^2 n_\mathrm{H} c}  \propto  {L}/{r^{2}n_\mathrm{H}}
\end{equation}

 \citep{netzer13}, where $Q(H)$ is the number of hydrogen ionizing photons. $U$ has to be approximately constant, otherwise we would observe a significant change in the spectral appearance.  
 


\end{enumerate}
 
 \subsection{The virial luminosity equation}

Taking the three constraints into account, the virial luminosity equation derived by \citet{marzianisulentic14} can be written in the form: 
\begin{eqnarray}\label{eq:vir}
L(\rm FWHM) & = & {\mathcal L_{\rm 0}} \cdot  (\rm FWHM)^{4}_{1000} \\ \nonumber 
 & = & 7.88 \cdot 10^{44} \left(\frac{L}{L_\mathrm{Edd}}\right)_{,1}^{2} \cdot   
 \frac{ \kappa_\mathrm{i,0.5}f_\mathrm{S,2}^2}{h \bar{\nu}_\mathrm{i,100 eV}} \frac{1}{(n_\mathrm{H}U)_{10^{9.6}}} (\rm FWHM)^{4}_{1000}\, {\rm erg  \, s}^{-1}
\end{eqnarray}
where the energy value has been normalized to 100 eV ($ \bar{\nu}_\mathrm{ i,100eV} \approx 2.42 \cdot 10^{16}$ Hz), $\kappa_\mathrm{i,0.5}$\ is the fraction of bolometric luminosity belonging to the ionizing continuum { scaled to 0.5},  the product ($n_{\rm H}U$) has been scaled to the typical value $10^{9.6} $cm$^{-3}$ \citep{padovanirafanelli88, matsuokaetal08,negreteetal12},  and the FWHM of the \hb\ BC  is expressed in units of1000 \kms. The $f_\mathrm{S}$ is scaled to the value 2 following the determination of \citet{collinetal06}.  The FWHM of \hb\ broad component and of \aliii\ are hereafter adopted as a virial broadening estimator. Eq. \ref{eq:vir}  implies that, by a simple measurement of the FWHM of a LIL, we can derive a $z-$independent estimate of the accretion luminosity  \citep[][c.f. \citealt{teerikorpi11}]{marzianisulentic14}. 

{ Recent works involved similar ideas concerning the possibility of ``virial luminosity" estimates, and confirm  that extremely accreting quasars could provide suitable distance indicators because their emission properties appear to be stable with their luminosity scaling with black hole mass at a fixed ratio \citep{wangetal13,lafrancaetal14,wangetal14b}.}  Eq. \ref{eq:vir} can be applied  to xA quasars only (\lledd $\sim$ 1). { However, it can be applied to {\em all xA} distributed   over a wide range of luminosity and redshift, where conventional cosmological distance incdicators are not available \citep[see][ for reviews concerning distance indicators beyond $z \gtrsim 1$]{donofrioburigana09,hook13,czernyetal18}.}

The virial luminosity equation is analogous to the Tully-Fisher and the early formulation of the Faber Jackson laws for early- type galaxies \citep{faberjackson76,tullyfisher77}. We note in passing that galaxies and even clusters of galaxies are virialized systems that show an overall consistency with a law $L\propto \sigma^\mathrm{n}$, where $\sigma$ is a measurement of the velocity dispersion and $n \sim 2 - 4$. Both the   \citet{faberjackson76} and \citet{tullyfisher77} found some application to the determination of the distance scale, especially at low-$z$ \citep[e.g., ][see \citealt{czernyetal18} for a recent review]{dressleretal87}. Distance measurements to early-type galaxies can be obtained by applying the relation of the fundamental plane of galaxies \citep[e.g.,][]{saulderetal19} which consider the relation between $\sigma$, structure and luminosity. Galaxies are more complex than quasars due to their non-homology \citep{donofrioetal17}. We hope to have minimized structural and physical condition differences by restricting the selection to xA quasars. { The factor $ {\mathcal L_{\rm 0}}$\ is most likely affected by intrinsic variations in the spectral energy distribution (SED) of  xA sources, as well as by anisotropic emission expected from a thick accretion disk \citep{wangetal14a}. The issue will be briefly discussed in \S \ref{sec:discuss}. }

\subsection{Orientation-dependent virial luminosity equation}
\label{angle}

The effect of orientation can be quantified  by assuming that the line broadening is due to an isotropic component + a flattened component whose velocity field projection along the line of sight is $\propto 1/\sin \theta$: 
\begin{equation}
\delta v^{2}_\mathrm{obs} = \frac{\delta v_\mathrm{iso}^{2}}{3} + \delta v_\mathrm{K}^{2} \sin^{2}\theta.
\label{eq:v}
\end{equation}

The structure factor  in the Eq. \ref{eq:vir} is set to $f_{\rm S} = 2$ since \citet{collinetal06} derived  values for $f_\mathrm{S} \approx$  2.1 for Pop. A, with a substantial scatter.   If we considered a flattened distribution of clouds with an isotropic $\delta v_{\rm iso}$ and a  velocity component associated with a rotating  flat disk $\delta v_{\rm K}$, the structure factor appearing in Eq. \ref{eq:vir}  can be written as 

\begin{equation}
f_{\rm S}  = \frac{1}{4 \left[ \frac{1}{3}\left(\frac{\delta v_{\rm iso}}{\delta v_{\rm K}}\right)^{2} + \sin^{2} \theta  \right]   } \label{eq:fs}
\end{equation}

which can reach values $\gtrsim 1$\ if $\kappa = \frac{\delta v_{\rm iso}}{\delta v_{\rm K}} \ll 1$, and if $\theta$ is also small ($\lesssim 30$ deg). The assumption $f_{\rm S} = 2$ implies that we are seeing a highly flattened system (if all parameters in Eq. \ref{eq:vir} are set to their appropriate values); an isotropic velocity field would yield $f_{\rm S} = 0.75$.   


The virial luminosity equation may  be rewritten in the form:
\begin{eqnarray}
L({\rm FWHM}) &=&    {{\mathcal L_{\rm 0}}}  f^{-2}_{\rm S} (\delta v_{\rm K})^{4}   \\ \nonumber
& = &  {  4^{2}   {\mathcal L^{\bullet}_{\rm 0}}  (\delta v_{\rm K})^{4}   }{\left[ \frac{\kappa^{2}}{3} + \sin^{2} \theta \right]}^{2} = 4^{2}  L_{\rm vir} {\left[ \frac{\kappa^{2}}{3} + \sin^{2} \theta \right]}^{2}, \\ \nonumber
 \end{eqnarray}
where $L_{\rm vir}$ is the true virial  luminosity (which implies $f_{\rm S} = 1$) with   ${\mathcal L^{\bullet}_{\rm 0} }= \frac{1}{4}{\mathcal L_{\rm 0}}$, since  ${\mathcal L_{\rm 0}}$ was scaled to $f_{\rm S} = 2$.



\begin{figure}[h!]
\begin{center}
\includegraphics[width=13cm]{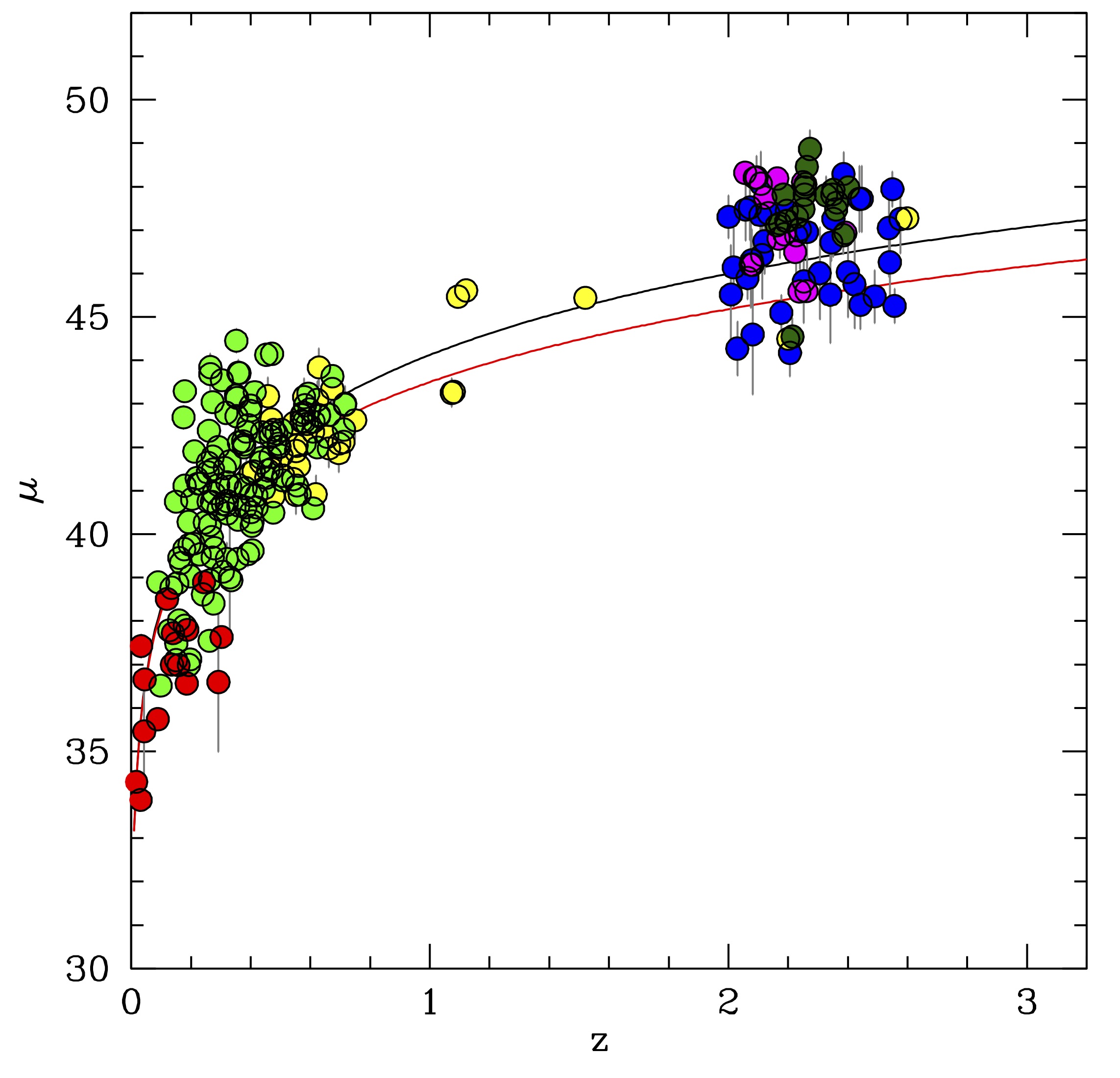}
\end{center}
\caption{Hubble diagram (distance modulus $\mu$ vs  redshift $z$) for several samples of xA quasars using virial luminosity distances from the FWHM of \hb\ and \aliii.  Data points are color-coded according to sample and line employed:  blue   (\textcolor[rgb]{0,0,1}{$\bullet$}): 
 \citet{marzianisulentic14} \aliii; yellow (\textcolor{yellow}{$\bullet$}):   \citet{marzianisulentic14} \hb; green (\textcolor[rgb]{0,1,0}{$\bullet$}): \citet{negreteetal18} \hb; red (\textcolor[rgb]{1,0,0}{$\bullet$}): \citet{duetal16a} \hb; dark green   (\textcolor[rgb]{0,0.3,0}{$\bullet$}): Sniegowska et al. in preparation; magenta (\textcolor{magenta}{$\bullet$}): \citet{martinez-aldamaetal18}. The continuous lines trace the prediction for concordance $\Lambda$CDM (black) the Einstein-de Sitter (red) cosmology.  
 \label{fig:hub}}
\end{figure}


\section{The Hubble diagram for quasars}
\label{cosmo}

\subsection{The basic}
\label{sec:hubble}
Following  \citet{weinberg72,weinberg08} the luminosity distance $d_{\rm L}$ can be written as:

\begin{eqnarray}
d_L(z;\Omega_M,\Omega_\Lambda, H_0)  & =&
\frac{d_{\rm H}(1 + z)}{  \sqrt{|\kappa | }} \cdot \; \; \; \\
& & \hspace{-3cm} {  S}\! \left (
 \sqrt{|{\kappa}| } \int_0^{z} \left [(1+z^\prime)^2(1+\Omega_M z^\prime)-
  z^\prime (2+z^\prime ) \Omega_\Lambda \right]^{-\frac{1}{2}} dz^\prime
 \right ),\nonumber
\end{eqnarray}
where $\Omega_M$ \ is  the energy density associated to  matter and   $\Omega_\Lambda$\  the energy density associated to  $\Lambda$, and 
$d_{\rm H}\equiv\frac{c}{H_0} = 3000\,h^{-1}~{\rm Mpc}= 9.26\times10^{25}\,h^{-1}~{\rm m} $, with $\log\ d_\mathrm{H} \approx 28.12$\ [cm].  

If $\Omega_M + \Omega_\Lambda > 1$, ${ S}(x)$ is defined as
$\sin(x)$ and $\kappa = 1 - \Omega_M - \Omega_\Lambda $; for $\Omega_M +
\Omega_\Lambda < 1$, ${ S}(x) = \sinh(x)$ and $\kappa$ as above; and
for $\Omega_M + \Omega_\Lambda = 1$, ${ S}(x) = x$ and $\kappa =1$. 

The equation of \citet{perlmutteretal97} has been obtained by   posing $\Omega_k +\Omega_{\Lambda}+ \Omega_\mathrm{M} = 1$, where $\Omega_k$\ is the energy density associated with the   curvature of space-time. The luminosity distance $d_{\rm L}  = d_{\rm C} \cdot (1+z)$ can be computed  in practice from:\begin{equation}
d_{\rm C} = d_{\rm H}\,\int_0^z\frac{dz'}{E(z')}
\end{equation} with 
\begin{equation}
\label{eq:dh}
d_{\rm H}\equiv\frac{c}{H_0}
= 3000\,h^{-1}~{\rm Mpc}= 9.26\times10^{25}\,h^{-1}~{\rm m} 
\end{equation}
with $\log\ d_\mathrm{H} \approx 28.12$ [cm] and 
\begin{equation}
\label{eq:ez}
E(z)\equiv\sqrt{\Omega_{\rm M}\,(1+z)^3+\Omega_k\,(1+z)^2+\Omega_{\Lambda}}
\end{equation}

The distance modulus   $\mu$ is defined by
\begin{equation} \label{eq:mu2}	
 \mu   \equiv   5\,\log \left(\frac{d_{\rm L}(H_{0}, \Omega_{\rm M},\Omega_k,\Omega_{\Lambda},  z)}{10~{\rm pc}}\right),  
 \end{equation}

that is $\mu = \log d_\mathrm{L} (H_{0}, \Omega_{\rm M},\Omega_k,\Omega_{\Lambda},  z) + 5$,  where the constant +5 becomes +25 if distances are expressed in Mpc \citep{lang80}.

The distance modulus $\mu$ computed from the virial equation yielding $L(\delta v)$ can be written as:
\begin{equation}
\mu  = 2.5 [\log  L(\delta v) - BC] - 2.5 \log  (f_{\lambda} \lambda)  -2.5 \log (4 \pi \delta_\mathrm{10pc}^{2}) + 5 \cdot \log (1+z)\label{eq:mu0}
\end{equation}

where the constant  $-2.5 \log (4 \pi \delta_\mathrm{10pc}^{2})$ =-100.19, with $\delta_\mathrm{10pc} \approx 3.08 \cdot 10^{19}$\, the distance of 10pc expressed in cm. The  $f_{\lambda} \lambda$\ can be the flux at 5100 \AA\ for the \hb\ sample, or  the flux at 1700 \AA\ if the $\delta v$=FWHM comes from the \aliii\ and \siiii\ lines. The bolometric correction $BC$\ has also to be changed accordingly.

The $\delta \mu = \mu_\mathrm{vir} - \mu(z;\Omega_M,\Omega_\Lambda, H_0)$ can be written as:

\begin{equation}
\delta\mu  = 2.5 [\log  L(\delta v) - BC] - 2.5 \log  (f_{\lambda} \lambda)    + 5 \cdot \log (1+z) -   5\log d_{\rm L}(H_{0}, \Omega_{\rm M},\Omega_k,\Omega_{\Lambda},  z)
\end{equation}

Following the notation of \citet{marzianisulentic14}  who wrote the comoving distance as $d = c/H_{0} \mathcal{F}(\Omega_{\rm M},\Omega_k,\Omega_{\Lambda},  z)$, we derive:\hfill
\begin{eqnarray}\label{eq:mu1}
\mu & \approx  &5  \cdot \left( \log \frac{c}{H_{0}} + \log \mathcal{F}(\Omega_{\rm M},\Omega_k,\Omega_{\Lambda},  z)\right) \\ \nonumber
&& - 5 \cdot \log \delta_\mathrm{10pc} + 5 \cdot \log (1+z)
\end{eqnarray}

If we equate  Equations  \ref{eq:mu0} and \ref{eq:mu1} above, we obtain: 
 
 \begin{eqnarray}
2.5 [\log  L(\delta v) - BC] - 2.5 \log  (f_{\lambda} \lambda) - 5 \log \delta_\mathrm{10pc}&& \\ \nonumber
 - 2.5 \log 4\pi   + 5 \cdot \log (1+z) & = &\\ \nonumber
5 \cdot \left( \log \frac{c}{H_0} + \log \mathcal{F}(\Omega_{\rm M},\Omega_k,\Omega_{\Lambda},  z)\right)\\ \nonumber
 - 5 \cdot \log \delta_\mathrm{10pc} + 5 \cdot \log (1+z).\\ \nonumber
\end{eqnarray}

Simplifying:



\begin{equation}
  { [L(\delta v) - BC]}     = 
 4\pi \left(  \frac{c}{H_0} {\mathcal{F}}(\Omega_{\rm M},\Omega_k,\Omega_{\Lambda},  z)\right)^{2}{ (f_{\lambda} \lambda),  } 
\end{equation}

we  recover the expression used by \citet{marzianisulentic14} who used only luminosity and did not give any expression for  $\mu$. Note that ${ (f_{\lambda} \lambda)  }$  refers to the rest frame fluxes (the term $(1+z)^{2}$\ appears in both side of the equations, for the virial luminosity and for the distance modulus where the luminosity distance was used). 

\subsection{Results}

The Hubble diagram of Fig. \ref{fig:hub} includes  sources from \citet{marzianisulentic14}, with both \hb\ and \aliii, and is supplemented by   SDSS low-luminosity xA sources at low $z$\ covering the H$\beta$\ range \citep{negreteetal18},  by   new high-$z$ quasar observations  at redshift $2 \lesssim z < \lesssim 3$\ that were observed with the OSIRIS camera at the GTC, covering the $\lambda$1900 \AA\ blend and hence \aliii\ \citep{martinez-aldamaetal18}.  An additional sample at very low-$z$\ is the one of the xA sources considered for the \citet{duetal16a} fundamental plane. Preliminary results of the \aliii\ measurements for $\approx$ 15 quasars (Sniegowska et al. 2019, in preparation) are also shown.  The value of  $\log$ ${\mathcal L_{0}} \approx$  45.06 was used in the  computation of the distance modulus $\mu$\ for  this work.  The plots in Fig. \ref{fig:hub} involves a total of 253  sources and indicate a scatter $\delta \mu \approx $ 1.2 mag. { A similar result was shown with 169 xA quasars by \citet{marzianietal17}. }
The plot of  Fig. \ref{fig:hub}  shows that even our heterogeneous sample  joining \hb\ and \aliii\ based sub-samples already rules out extreme Universes such as the Einstein-de Sitter Universe ($\Omega_\mathrm{M}=1, \Omega_\Lambda = 0$, { red line in Fig. \ref{fig:hub}}), which under-predicts $\mu$ values at high $z$\ by $\approx 1 $ mag {  with respect to concordance. The inclusion of the \aliii\ has been instrumental to the extension of the Hubble diagram to redshifts $\gtrsim 1$. Observations of \hb\ at redshift $\gtrsim$ 1 are available for several hundreds of quasars \citep[e.g.,][]{coatmanetal16,shen16}; this means that only a few tens of spectra may be available from literature, if the observations are restricted to xA source and high S/N is requested. An additional virial broadening estimator is needed. The \aliii\  line can be measured from SDSS survey up to $z \gtrsim 4$. Fig. \ref{fig:hbal} shows preliminary results for a sample of Population A sources. The evidence collected so far suggests that the correlation should remain valid for {\em most} xA sources (a few notable exceptions are known; for example {HE0359-3959}, \citealt{martinez-aldamaetal17}). The suitability of \aliii\ as a virial broadening estimator will be further explored in eventual works.}


The Hubble diagram for quasars restricted to the 92 sources of \citet{marzianisulentic14}  was also consistent with concordance $\Lambda$CDM provided some constraints on  $\Omega_\mathrm{M}:  0.19^{+0.16}_{-0.08}$. The  redshift    range 2 - 3 is highly sensitive to $\Omega_\mathrm{M}$. The new sample is consistent with the old one and allows to set more stringent limits to  $\Omega_\mathrm{M} \approx$ 0.30$^{+0.06}_{-0.06}$\ via a standard Bayesian least square fit, employing a sample with no restriction on data (i.e., including low-S/N \aliii\ spectra and a few quasars at $z$\ in the range 2.6 —3 not shown in Fig. \ref{fig:hub}).  

\section{Discussion}
\label{sec:discuss}

{ In this paper we have described the method developed by \citet{marzianisulentic14} in more detail, presented an expanded Hubble diagram with more objects with respect to the one presented by \citet{marzianietal17}, as well as an updated (but still preliminary estimate) of $\Omega_\mathrm{M}$ and the Hubble diagram with concordance and Einstein-de Sitter cosmologies, just to illustrate the sensitivity of the method to cosmological parameter changes. } { It should be also noted that} a comparison between the constraints set by the supernova photometric survey described by \citet{campbelletal13}  and a mock sample of 400 quasars with rms = 0.3 uniformly covering the $z$ range $0.1-3$ that assumes concordance cosmology (shaded contours in Fig. 4 of \citealt{marzianisulentic14a}) emphasizes the  potential ability of the quasar sample to better constrain $\Omega_\mathrm{M}$. 

An elementary error budget suggests that the main uncertainty is associated with FWHM measurement errors  and with uncertainty in the  structure factor \citep{marzianisulentic14,negreteetal18}. The broad profile of both LILs and HILs in each quasar spectrum can be modeled  by considering two main components, the \hb\ BC and the blueshifted component BLUE (\S \ref{ms}).    Results shown in Fig. \ref{fig:hub}  { refer to \hb\ BC profiles whose FWHM has been corrected for the effect of BLUE via  a multi-component fit, as described in \citet{negreteetal18}.}   The \hb\ BLUE affects however the \hb\ FWHM only at $\approx $ 10\% level, with an effect that should be $\lesssim $ 0.3 dex. Vetting the sample on strongest \feii\ and weakest \oiiiopt\ emitters reduces the scatter but only slightly \citep{negreteetal18}. 

\citet{negreteetal18} have convincingly shown that  orientation effects are at the origin of a large fraction of the scatter   if lines are emitted in a flattened system ($\kappa \lesssim 0.2$). The difference between virial and concordance luminosity estimates can be zeroed if the structure factor is assumed to be dependent on angle $\theta$\ in the form given by Eq. \ref{eq:fs}. The resulting distribution of $\theta$ values ranges from 0 to 50 with a peak around $\theta \approx 18$. { The range of viewing angles in the sample of \citet{negreteetal18} is actually modest.  The effect of anisotropy from a geometrically and optically  thick disk on the SED is not  extremely strong within $\theta \lesssim 30$, approximately around 0.2 dex, following the models of anisotropic emission of \citet{wangetal14a}. The  actual effect on virial luminosity estimates could be even less than that, as the dispersion in the distribution of $\theta$\ in the sample of \citet{negreteetal18} is just 7 degrees, which  implies that most  sources are seen in a viewing angle range between 10 and 30 degrees. Fig. 4 of \citet{wangetal14a} indicates that different $\theta$\ should lead to an increase of the dispersion of a few 0.01 dex around the luminosity expected for 20 degrees at optical wavelengths. This will mainly affect the continuum $\lambda f_{\lambda} $\ that we receive  entering in Eq. \ref{eq:mu0}. }

{ The presence of a thick disk implies self-shadowing effects, and that the line emitting gas might not be exposed to the same ionizing continuum that is seen by the observer. This second effect related to the disk anisotropy is especially severe if the line emitting gas is located close to the equatorial plane of the disk: in this case, following \citet{wangetal14a}, we expect a drastic reduction of the number of ionizing photons which would affect ${\mathcal L_{\rm 0}}$. Precisely quantifying  the effect on line emission is not trivial, as it depends on the distribution of the line emitting gas around the equatorial plane of the disk, in terms of both angular and radial distance \citep{duwang19},  the black hole mass  as well as the accretion disk model. Differences in the accretion disk SED are also expected from object to object because of different black hole spins \citep{wangetal14}. The  spin and the anisotropy effects on ${\mathcal L_{\rm 0}}$ should be considered in an eventual work, but  there is a basic point that we want to remark about the usability of Eq. \ref{eq:vir} for cosmology. 


The SED parameters  entering in the relation for the virial luminosity are $\bar{\nu}_\mathrm{i}$, $\kappa_{\mathrm{i}}$, and more indirectly, $n_\mathrm{H}U$. The ionizing photon flux $n_\mathrm{H}U$\ shows remarkably small changes along the MS (i.e., even without restricting the attention to xA sources), as shown by the various estimates reported in Table 4 of \citet{negreteetal12}: it cannot change by a large factor, otherwise the emission line spectrum would change dramatically \citep{pandaetal18,pandaetal19}.    The other SED-related parameters  appear  as  $\kappa_{\mathrm{i}}/\bar{\nu}_\mathrm{i}$,  and are qualitatively affected in the same way by a change of the ionizing continuum.  Explorative computations using CLOUDY \citep{ferlandetal17} indicate that the ratio $\kappa_{\mathrm{i}}/\bar{\nu}_\mathrm{i}$ decreases by a factor $\approx 1.2$  for a change of $\theta$ from $\theta \lesssim 20 - 30$ to $\theta \sim 50 - 70$. This condition could be appropriate if the ``walls'' of the thick disk are illuminating the low-ionization part of the BLR.   Object-by-object differences may add scatter in ${\mathcal L_{\rm 0}}$, therefore contribution to the dispersion observed in the Hubble diagram of Fig. \ref{fig:hub}.  

The issue of anisotropy and of other factors affecting the ${\mathcal L_{\rm 0}}$ term multiplying the line width in Eq. \ref{eq:vir} brings to the attention the possibility of structural changes with redshift and luminosity although it is at least conceivable that the accretion mode at high rates is giving rise to a stable and reproducible structure. This consideration is supported by the stability of xA sources at least in the optical domain \citep[][]{duetal18}, as shown by the small amplitude of continuum variations detected in dedicated reverberation mapping campaigns, and by the similarity of low-ionization emission line profiles over a wide range of redshift and luminosity.  }

{ Summing up, we can say that the proper calibration of the data for cosmology will need large samples of quasar spectra covering the rest frame optical and UV, at high S/N, over a wide range of redshift. This is a {\em conditio sine qua non} to test for systematic sources of errors and hence for the effective  exploitation of  quasar data for cosmology applying the virial luminosity equation at $z \gtrsim 1$.} 



\section{Conclusion}
 
The quasar MS has allowed us to isolate sources which are the highest radiators among quasars. This is already an important feat, as xAs can then be selected  by the application of simple criteria based on optical and UV emission line ratios. xA quasars show a relatively high prevalence (10\%) and are easily recognizable in the redshift range 1 -- 5.  { The present work illustrated several key aspects of the method, and provided the virial luminosity equation dependent from the viewing angle in an explicit form (a result of \citealt{negreteetal18}). However, several additional caveats need to addressed in subsequent work: the reliability of \aliii\ as a virial broadening estimator and, generally speaking, the inter-calibration of the data obtained in the rest frame UV and optical domain, as well as the effect of anisotropic continuum emission expected for xA sources. } xA  might be suitable as Eddington standard candles especially if orientation { and systematic effects as a function of redshift and luminosity can be identified} and accounted for.

\section*{Author Contributions}

PM  wrote the paper with the assistance of DD; PM and JAdD did the Hubble diagram analysis and JAdD contributed the $\Omega_\mathrm{M}$\ estimate. All other authors (part of the {\em extreme team}) sent comments and/or contributed to some extent during the development of the project. 

\section*{Funding}

 \section*{Acknowledgments}

DD and AN acknowledge support from grants  PAPIIT, UNAM 113719, and CONACyT221398. PM and MDO acknowledge funding from the INAF PRIN-SKA 2017 program 1.05.01.88.04. PM also acknowledges the Programa de Estancias de Investigaci\'on (PREI) No. DGAP/DFA/2192/2018 of UNAM where this work was advanced, and  the {\em Hypatia of Alexandria} visiting grant SO-IAA (SEV-2017-0709).  AdO acknowledge financial support from the Spanish Ministry of Economy and Competitiveness through grant AYA2016-76682-C3-1-P and from the State Agency for Research of the Spanish MCIU through the ``Center of Excellence Severo Ochoa" award for the Instituto de Astrof\'{\i}sica de Andaluc\'{\i}a (SEV-2017-0709). 
 

\bibliographystyle{frontiersinSCNS_ENG_HUMS}  







\end{document}